\documentclass[aps,prl,twocolumn,superscriptaddress,showpacs]{revtex4}

\usepackage{graphicx}
\usepackage{amssymb}
\usepackage{amsmath}

\begin{document}


\title{Observation of single collisionally cooled trapped ions in a buffer gas}



\author{M.~Green}
\author{J.~Wodin}
\author{R.~DeVoe}
\author{P.~Fierlinger}
\author{B.~Flatt}
\author{G.~Gratta}
\author{F.~LePort}
\author{M.~Montero~D\'{i}ez}
\author{R.~Neilson}
\author{K.~O'Sullivan}
\author{A.~Pocar}
\author{S.~Waldman}\thanks{Now at Caltech, Pasadena CA, USA}\affiliation{Physics Department, Stanford University, Stanford CA, USA}

\author{D.S.~Leonard}
\author{A.~Piepke}\affiliation{Department of Physics and Astronomy, University of Alabama, Tuscaloosa AL, USA}

\author{C.~Hargrove}
\author{D.~Sinclair}
\author{V.~Strickland}\affiliation{Physics Department, Carleton University, Ottawa ON, Canada}

\author{W.~Fairbank Jr.}
\author{K.~Hall}
\author{B.~Mong}\affiliation{Physics Department, Colorado State University, Fort Collins CO, USA}

\author{M.~Moe}\affiliation{Physics Department, University of California, Irvine CA, USA}

\author{J.~Farine}
\author{D.~Hallman}
\author{C.~Virtue}\affiliation{Physics Department, Laurentian University, Sudbury ON, Canada}

\author{E.~Baussan}
\author{Y.~Martin} 
\author{D.~Schenker}
\author{J.-L.~Vuilleumier}
\author{J.-M.~Vuilleumier}
\author{P.~Weber}\affiliation{Institut de Physique, Universit\'{e} de Neuchatel, Neuchatel, Switzerland}

\author{M.~Breidenbach}
\author{R.~Conley}
\author{C.~Hall}\thanks{Now at University of Maryland, College Park MD, USA}
\author{J.~Hodgson}
\author{D.~Mackay}
\author{A.~Odian}
\author{C.Y.~Prescott}
\author{P.C.~Rowson}
\author{K.~Skarpaas}
\author{K.~Wamba}\affiliation{Stanford Linear Accelerator Center, Menlo Park CA, USA}


\date{\today}

\begin{abstract}
Individual Ba ions are trapped in a gas-filled linear ion trap and observed with a high signal-to-noise ratio by resonance fluorescence.  Single-ion storage times of $\sim$5~min ($\sim$1~min) are achieved using He (Ar) as a buffer gas at pressures in the range $8\times10^{-5} -  4\times10^{-3}$~torr.   Trap dynamics in buffer gases are experimentally studied in the simple case of single ions.  In particular, the cooling effects of light gases such as He and Ar and the destabilizing properties of heavier gases such as Xe are studied.   A simple model is offered to explain the observed phenomenology.
\end{abstract}

\pacs{52.20.Hv, 39.25.+k, 29.30.EP, 23.40.-s}

\maketitle


\section{Introduction}

The discovery of ion trapping \cite{Paul,Post} followed by pioneering experiments on single trapped ions in vacuum \cite{Neuhauser} has opened the door to a wealth of physics, such as laser cooling \cite{Itano,Wineland}, quantum computation \cite{Nielsen}, improved atomic frequency standards \cite{Oskay}, mass spectrometry \cite{Douglas}, and measurements of metastable atomic lifetimes \cite{Yu}.  Traps are usually operated in ultra-high vacuum (UHV, $\lesssim10^{-10}$~torr), and laser cooling techniques are utilized for precision measurements on a single confined ion.  Buffer gas filled RF Paul traps, on the other hand, use collisions to thermalize ions with the buffer gas.  These devices have proven useful, in particular, for mass spectrometry on ion clouds, as well as cooling stages in heavy isotope accelerators \cite{Kellerbauer, Herfurth}. Buffer gas cooling is effective at cooling many ions at a time, and charged species with many degrees of freedom such as ionized biological molecules \cite{March}.  The observation of individual buffer-gas-cooled ions via resonance fluorescence is challenging, primarily due to Doppler broadening of the ion's natural linewidths.  This limits the sensitivity of the technique as a tool to detect trace amounts of ions.  Furthermore, increasing the number of trapped ions to obtain more fluorescence makes it difficult to study ion dynamics and their interactions with the buffer gas, since ion-ion interactions produce non-negligible effects.

In this work, single Ba ions are trapped and continuously cooled by collisions with a room temperature buffer gas (He or Ar) at pressures between $8\times10^{-5}$ and $4\times 10^{-3}$~torr in a segmented quadrupole linear RF Paul trap.  The 6S$_{1/2}$~$\leftrightarrow$~6P$_{1/2}$ and 6P$_{1/2}$~$\leftrightarrow$~5D$_{3/2}$ transitions are driven with CW diode lasers, producing resonance fluorescence at 493~nm and 650~nm \cite{Neuhauser}.  Individual ions are stored for $\sim$50~s to $\sim$500~s and imaged.  Effective ``Doppler'' ion temperatures are measured from the broadening of the 6P$_{1/2}$~$\leftrightarrow$~5D$_{3/2}$ transition.   Also, single-ion unloading rates are measured at different pressures of He, Ar and He/Xe mixtures, providing insight into the cooling behavior of light gases and the de-stabilizing properties of heavy ones.   A simple collisional unloading model is found to be able to describe the data. 

\begin{figure}
\includegraphics[width=3.25in]{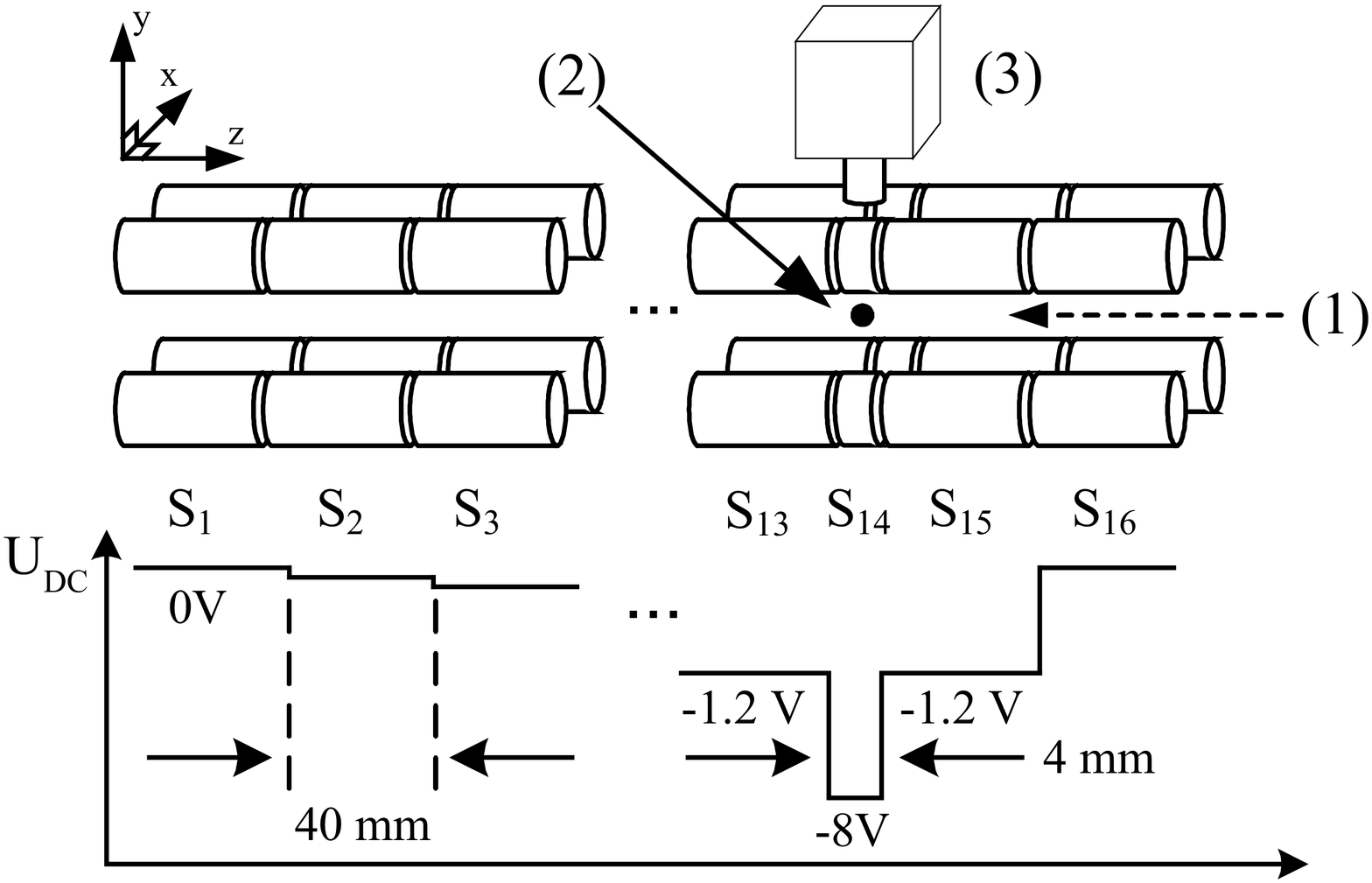}
\caption{Schematic drawing of the buffer-gas filled linear RF Paul trap used in this work.  Ions are created in S$_3$ via electron impact ionization, and collisionally cooled by the buffer gas into the potential well at S$_{14}$.  $U_{DC}$ is the potential applied to each segment.  The ion is identified via resonance fluorescence at 493 nm. (1) Spectroscopy lasers, (2) Individual Ba ion, (3) EMCCD camera. \label{fig:IonTrap}}
\end{figure}

Linear RF Paul traps confine ions radially in a quadrupole RF field applied across electrodes placed symmetrically about a central axis \cite{Paul1990}.  The trap used in this experiment is shown schematically in fig. \ref{fig:IonTrap}.  A three dimensional potential well is obtained at segment 14 (S$_{14}$) by creating a DC potential minimum along the z-axis of the trap, in addition to the radial RF pseudopotential minimum.    

The confinement of an externally injected ion of charge $e$ requires a kinetic energy dissipation mechanism, such that the ion is cooled to an energy less than both $D_r$ and $D_z$ while at the trap minimum.   One such energy dissipation mechanism is laser cooling, which requires UHV conditions in addition to cycling transitions used for cooling.  Energy dissipation in the system described here, in contrast, is achieved by collisions with a buffer gas.  The ion comes into thermal equilibrium with the buffer gas, settling into the well at S$_{14}$ with energy $\sim$0.03~eV for a gas at 293~K.   Over the range of pressures relevant here, Ba ions undergo 1-100 collisions with the buffer gas before reaching S$_{14}$ from the injection region at S$_{3}$.

The details of collisional cooling have been investigated analytically \cite{Dehmelt,Devoe,Moriwaki1992} and numerically \cite{Herfurth,Kim,Devoe,Waldman} for $m_I\gg m_B$, $m_I\sim m_B$, and $m_I\ll m_B$, where $m_I$ and $m_B$ are the masses of the ion and buffer gas atoms.  Consider an ion with initial kinetic energy $E_I$, interacting with a buffer gas at temperature $T_B<E_I/k_B$, where $k_B$ is Boltzmann's constant.   If $m_I\gg m_B$, each collision perturbs the ion's momentum by a small amount, bringing the ion into thermal equilibrium with the buffer gas after many collisions.  Data relevant to this situation are presented in the following section, where Ba ions ($m_I\approx137$ amu) are trapped in He ($m_B\approx$ 4 amu) and Ar ($m_B\approx$ 40 amu) buffer gases.   If $m_I\sim m_B$, the ion is cooled to the temperature of the buffer gas with far fewer collisions, at the expense of a higher probability per collision of unloading from the trap.  This unloading is due to RF-heating, whereby a large momentum change of the ion per collision can cause the ion to absorb energy from the RF field.  Buffer gases composed of He/Xe mixtures ($m_{\mathrm{Xe}}\approx$ 131 amu) allow for the study of deconfining effects in this case, while retaining trapping times long enough to comfortably observe a single ion.   In the third case, $m_I\ll m_B$, the ion is not cooled, leading to unloading.  This case, treated in \cite{Moriwaki}, is not discussed further here. 


\section{Experimental Setup}

The construction and operational details of the ion trap used here are described elsewhere \cite{Trap,Wodin}.  The trap is housed in an all-metal UHV tank, coupled to a buffer gas purification and injection system.  The trap is split into fifteen 40~mm segments (S$_{1}$-S$_{13}$, S$_{15}$-S$_{16}$ in fig. \ref{fig:IonTrap}), and one 4~mm segment (S$_{14}$) defining the center of the trapping region.  The total trap length is 604~mm, chosen to be long enough to thermalize highly energetic, externally injected ions in future experiments.

Each segment consists of four, 6~mm diameter cylindrical stainless steel electrodes.  Radial confinement is achieved by applying a 150~V$_\mathrm{pk}$ sinusoidal voltage at 1.2~MHz to two diagonal electrodes, spaced  11.3~mm apart center-to-center, while keeping the other two at RF ground. Longitudinal confinement is achieved by applying a DC potential, U$_\mathrm{DC}$, to each segment as shown in fig. \ref{fig:IonTrap}. This configuration, similar to that used in ion coolers at heavy-ion facilities \cite{Kellerbauer}, gives a radial trap depth $D_{r} = 3.8$~eV, and a longitudinal trap depth $D_{z}=6.8$~eV \cite{Herfurth,Wodin}.

External cavity diode lasers at 493~nm (frequency doubled 986~nm) and 650~nm, locked to the optogalvanic resonance of a Ba discharge lamp \cite{Badareu}, create resonance fluorescence from individual trapped ions.  The beams are directed along the longitudinal axis of the trap, as illustrated in fig. \ref{fig:IonTrap}.  The 493~nm and 650~nm laser powers are set to 65~$\mu$W and 200~$\mu$W, respectively.   The 493~nm fluorescence from individual Ba ions is imaged onto an electron multiplying CCD\footnote{Andor model iXon$^\mathrm{EM}$+} (EMCCD) via a 64~mm working-distance microscope outside the vacuum chamber.  A spherical mirror located inside the chamber doubles the acceptance of fluorescence photons.   The fluorescence intensity is obtained by integrating the EMCCD signals over a region of interest centered around the trapping minimum.  The total fluorescence detection efficiency is estimated to be $\sim10^{-2}$.

Detection of the fluorescence from a single ion in the conditions described here is challenging because of the Doppler broadening of the spectral lines and the larger volume of space occupied by the ions.  At 293~K, the Doppler broadened lineshapes are a convolution of the natural ones with the ion's velocity profile along the laser axis. The Doppler linewidth is
\begin{equation}
\Gamma_D=\left(\frac{\omega_0}{c}\right)\sqrt{\frac{8k_BT\log{2}}{m_I}}
\label{eqn:doppler}
\end{equation}
where $c$ is the speed of light, $\omega_0$ is the frequency of the 493~nm or 650~nm transition, and $T$ is the Maxwell-Boltzmann temperature of the ion's motion. The natural linewidths of the two transitions are $\Gamma_{493}/2\pi=15.2$~MHz and $\Gamma_{650}/2\pi=5.3$~MHz \cite{Janik}.  Doppler broadening reduces the fluorescence from each of these transitions by a factor $\Gamma_D/\Gamma_{493,650}$, or 42 and 92, respectively.  The ion's spatial distribution in the trap is determined by the gas temperature and the geometry of the potential well.  In this experiment, the ion is confined to a volume $\sim(500~\mu m)^3$.  Imaging over such a large area increases the amount of scattered light accepted by the detection optics.  The simultaneous optimization of the laser beam quality, fluorescence collection optics, trapping segment size, and potential well depth is therefore essential to observe individual ions.

The buffer gas is purified by a hot Zr alloy getter \footnote{SAES MonoTorr model PF3C3R1}, that nominally reduces the contamination from electronegative impurities to $\lesssim1$~ppb.  To avoid the buildup of impurities from outgassing, the purified gas flows continuously through the vacuum system and is removed by a dry $520~\ell$/s turbomolecular pump.  The pressure is controlled by a flowmeter which is servoed to a pressure gauge sampling the vacuum chamber near S$_{14}$.  Ba ions are created at S$_3$ (far from the potential well) via electron impact ionization of neutral Ba, emitted from a Ba-coated Ta oven placed just outside of the trap.  Collisional cooling to the trap minimum at S$_{14}$ is achieved in much less than 1~s.

\section{Results}

\begin{figure}
	\includegraphics[width=3.25in]{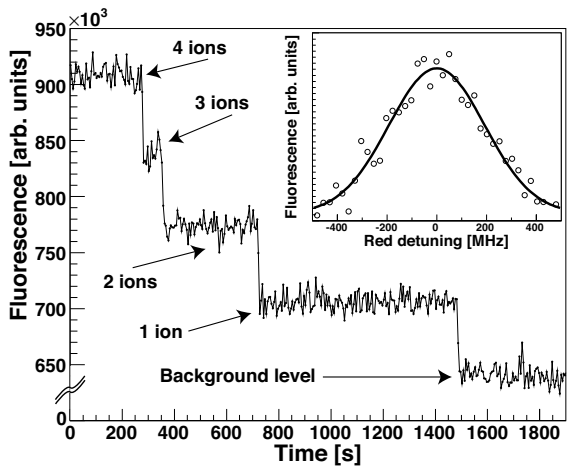}
	\caption{Intensity of the 493~nm fluorescence from single Ba ions trapped in S$_{14}$ as a function of time. The ions are continuously cooled by $8.6\times10^{-4}$~torr He at 293~K.  The horizontal axis in the inset is the frequency detuning of the light exciting the 6P$_{1/2}\leftrightarrow$~5D$_{3/2}$ transition.    The vertical axis is the fluorescence intensity.   Zero detuning corresponds to the peak of the fit used to calculate the ion's temperature.  \label{fig:Quantization}}
\end{figure}

In fig. \ref{fig:Quantization}, four ions are initially loaded into the trap at $8.6\times10^{-4}$~torr He, and the 493~nm fluorescence intensity in the region of interest is continuously observed over time.  Individual ions spontaneously eject from the trap, reducing the fluorescence intensity by quantized amounts.  Each data point in the time series corresponds to 5~s of integrated fluorescence by the EMCCD, in arbitrary units of EMCCD counts.   
Single ions are clearly identified with a signal to noise ratio of 10.6 (defined as the distance, in number of sigmas, between the 0-ion and 1-ion cases).  The ability to identify single ions in such an un-ambiguous way, and distiguish them from the background level, opens the possibility of collecting very clean data on the unloading properties of different gases, as required to understand trap dynamics in these conditions.   The signals from more than one ion, useful to establish single ion detection, are not used in the rest of this paper. 

The Maxwell-Boltzmann temperature of a single ion in He is measured by scanning the red laser frequency over the Doppler broadened 6P$_{1/2}\leftrightarrow$~5D$_{3/2}$ resonance (fig. \ref{fig:Quantization}, inset).  The 493~nm fluorescence intensity is monitored, while the blue frequency is kept fixed.  The ion's temperature is extracted by fitting the measured spectrum to a Voigt profile, taking into account the measured power broadening due to the lasers.  This profile assumes a Maxwellian velocity distribution along the laser axis.  The temperature does not show a significant correlation with buffer gas pressure between $5\times10^{-5}$ and $4\times10^{-3}$~torr of He.  A $\chi^2$ fit of the temperature values obtained at different pressures and laser powers provides an average temperature of 256$\pm$10~K.  However, the $\chi^2$/dof of the fit is 24/12, indicating that further systematic errors not included in our analysis are present.

\begin{figure}
	\includegraphics[width=3.25in]{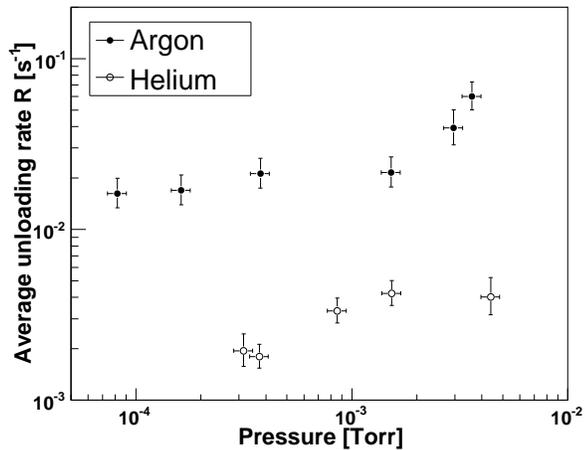}
	\caption{Average unloading rates of a single Ba ion in various buffer gases, as a function of pressure.\label{fig:Lifetimes}}
\end{figure}

The average single ion unloading rates in He and Ar are measured by loading a few ions into the trap.  Once a single ion is left, its residence time is measured.  This procedure is repeated multiple times at different pressures of He, Ar, and various He/Xe mixtures. Assuming the ion has a constant probability of unloading per unit time (or collision), the probability of an ion surviving for time $t$ in the trap is
\begin{equation}
\mathcal{P}_S(t)=e^{-Rt}
\end{equation}
where $R$ is the average unloading rate (inverse of the average storage time).  The average single ion unloading rates as a function of buffer gas pressure and  type are determined via an un-binned maximum likelihood method, using $\geq20$ independent residence time measurements for each pressure.  These results are shown in fig. \ref{fig:Lifetimes} for the case of pure light buffer gases.   The vertical errors in the figure are uncorrelated statistical fluctuations in the lifetimes.  The horizontal errors are the precision to which the absolute gas pressure is known near S$_{14}$.

Two qualitative features of the data are worth noting.  First, the average unloading rate increases with increasing buffer gas pressure.  This is consistent with an increase in the frequency of RF-heating collisions, or possibly, impurities present in the buffer gas.  Second, the unloading rate is lower for He than for Ar.  This is likely due to kinematic processes such as RF-heating, which depend on the buffer gas mass \cite{Devoe, Moriwaki}.

\begin{figure}
	\includegraphics[width=3.25in]{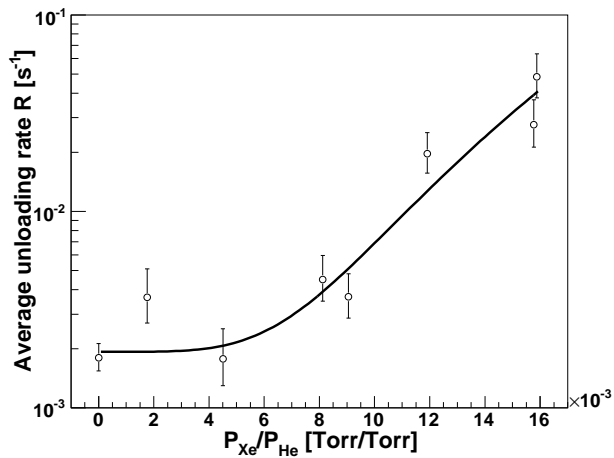}
	\caption{Measured unloading rate of a single ion stored in $3.7\times10^{-4}$~torr of He/Xe mixtures at different Xe concentrations.\label{fig:HeXeLifetimes}}
\end{figure}

The single ion unloading rate in various He/Xe mixtures is measured as a function of Xe concentration, as shown in fig. \ref{fig:HeXeLifetimes}.
In this case, the He acts as a continuous cooling mechanism, whereas collisions between the Ba ion and Xe atoms have a significant probability of ejecting the ion from the trap.  

A simple phenomenological model is fit to the data, based on the following argument.  The time intervals between Ba$^{+}$-He and Ba$^+$-Xe collisions are each exponentially distributed with rate constants $R_\mathrm{He}=C_\mathrm{He}P_{\mathrm{He}}$ and $R_\mathrm{Xe}=C_\mathrm{Xe}P_{\mathrm{Xe}}$, respectively, where $P_\mathrm{He,Xe}$ are the He and Xe buffer gas partial pressures.  Collision physics details are absorbed into the coefficients $C_\mathrm{He,Xe}$.  The probability that a Ba ion collides once with a Xe atom in a time $t$ is
\begin{equation}
\mathcal{P}_C^\mathrm{Xe}(t)=1-e^{-R_\mathrm{Xe}t}=1-e^{-C_{\mathrm{Xe}}P_{\mathrm{Xe}}t}
\label{eqn:OneCollision}
\end{equation}
As evident from the magnitude of the ejection rates in fig. \ref{fig:HeXeLifetimes}, compared to the Ba$^+$-Xe collision frequency ($\sim$~kHz), a single Ba$^+$-Xe collision is not sufficient, on average, to unload an ion.  Instead, unloading appears to require multiple consecutive Ba$^+$-Xe collisions on a timescale less than the Ba$^+$-He collision time, $R_\mathrm{He}^{-1}$.  The probability of $n$ Ba$^+$-Xe collisions during a time $t\leq R_\mathrm{He}^{-1}$ is
\begin{equation}
\prod_n\mathcal{P}_C^{\mathrm{Xe}}(t\leq R_\mathrm{He}^{-1})=\left[1-\exp{\left(-\frac{C_{\mathrm{Xe}}P_{\mathrm{Xe}}}{C_{\mathrm{He}}P_{\mathrm{He}}}\right)}\right]^n
\label{eqn:NCollisions}
\end{equation}
For $P_{\mathrm{Xe}}\ll P_{\mathrm{He}}$,
\begin{equation}
\prod_n\mathcal{P}_C^{\mathrm{Xe}}(t\leq R_\mathrm{He}^{-1})\approx\left(\frac{C_{\mathrm{He}}P_{\mathrm{Xe}}}{C_{\mathrm{Xe}}P_{\mathrm{He}}}\right)^n
\label{eqn:NCollisionExpansion}
\end{equation}
Hence, the rate of this process is proportional to the $n^\mathrm{th}$ power of the Xe concentration.  Using this model, the data in fig. \ref{fig:HeXeLifetimes} is fit to the three parameter function,
\begin{equation}
R\left(\frac{P_\mathrm{Xe}}{P_\mathrm{He}}\right)=C_0+C_1\left(\frac{P_\mathrm{Xe}}{P_\mathrm{He}}\right)^n
\label{eqn:fitf}
\end{equation}
where $C_0$ is the unloading rate due to He alone, $C_1$ represents a combination of the Ba$^+$-Xe and Ba$^+$-He collision physics constants, and $n$ is the average number of collisions required to unload an ion.  The fit yields $n=4.5\pm0.6$ collisions (and $C_0=(1.8\pm0.3)\times10^{-3}$, $C_1=30\pm4$), with a $\chi^2$/dof = 8.6/5.


\section{Conclusions}

The ability to perform clean measurements on single ions in the presence of different background gases 
may provide important insights with regard to the dynamics of these systems.   Indeed, in the past,
the dynamics of trapped ions with a buffer gas was experimentally studied in systems where additional
complications arise from ion-ion interactions and space-charge effects.    The single ion measurements
described here allow for more accurate tests of the theory. From a practical point of view, traps 
operated in the regime discussed here may help to improve the sensitivity of detecting very rare 
atomic species with suitable fluorescence transitions and line-widths in less-than-ideal situations, 
where some gas contamination is unavoidable.    In particular this technique is relevant for the 
development of a very large double-beta ($\beta\beta$) decay experiment in which individual 
$^{136}$Ba$^+$, produced in the $\beta\beta$ decay of $^{136}$Xe, need to be identified with 
high resolution spectroscopy in the presence of Xe contamination~\cite{Breidenbach,Danilov}. 

\section{Acknowledgments}
This work was supported, in part, by DoE grant FG03-90ER40569-A019 and by private funding from Stanford University.  We also gratefully acknowledge a substantial equipment donation from the IBM corporation, as well as very informative discussions with Guy Savard.

\bibliography{other}

\end{document}